# NONPARAMETRIC SPECTRAL ANALYSIS WITH APPLICATIONS TO SEIZURE CHARACTERIZATION USING EEG TIME SERIES

By Li Qin[1] and Yuedong Wang[2]

*Fred Hutchinson Cancer Research Center and University of California, Santa Barbara*

Understanding the seizure initiation process and its propagation pattern(s) is a critical task in epilepsy research. Characteristics of the pre-seizure electroencephalograms (EEGs) such as oscillating powers and high-frequency activities are believed to be indicative of the seizure onset and spread patterns. In this article, we analyze epileptic EEG time series using nonparametric spectral estimation methods to extract information on seizure-specific power and characteristic frequency [or frequency band(s)]. Because the EEGs may become nonstationary before seizure events, we develop methods for both stationary and local stationary processes. Based on penalized Whittle likelihood, we propose a direct generalized maximum likelihood (GML) and generalized approximate cross-validation (GACV) methods to estimate smoothing parameters in both smoothing spline spectrum estimation of a stationary process and smoothing spline ANOVA time-varying spectrum estimation of a locally stationary process. We also propose permutation methods to test if a locally stationary process is stationary. Extensive simulations indicate that the proposed direct methods, especially the direct GML, are stable and perform better than other existing methods. We apply the proposed methods to the intracranial electroencephalograms (IEEGs) of an epileptic patient to gain insights into the seizure generation process.

**1. Introduction.** Roughly 1% of the population in developed nations suffers from epilepsy. Of these about 30% have medically refractory epilepsy, where the most devastating feature is seizure. The only hope for relieving

Received May 2008; revised May 2008.
[1]Supported by the Career Development Funds from the Fred Hutchinson Cancer Research Center.
[2]Supported in part by an NSF Grant.

these patients from the disabling seizures is resective surgery, while the surgical success rate varies between less than 25% to 70% depending on how well the seizure initiation zone could be removed or the seizure propagation path could be disconnected [Schiller, Cascino and Sharbrough (1998)]. Therefore, a better understanding of the seizure initiation process and its propagation patterns is crucial to the success of a surgery. Epileptic patients usually undergo presurgical evaluations where they are monitored over time with EEGs recorded by electrodes placed at different locations of the brain called channels. The locations of the EEG channels are believed to give the best coverage of the epileptogenic zone, determined by the neurosurgeons. While EEGs from 128 or 256 channels are recorded, only a few of them cover the origin or critical path of the seizure generation. Therefore, the identification of the seizure focus and its spread patterns would improve clinical judgement on where to resect that would render the patients seizure free. Characteristics of the pre-seizure ("pre-ictal") EEGs including oscillating powers and high-frequency activities are believed to be indicative of the seizure onset and spread patterns. They behave differently from those characteristics of the baseline ("inter-ictal") EEGs [Winterhalder et al. (2003)].

Spectral analysis of EEG time series plays a central role in epilepsy research. Due to the large heterogeneity in pathology between patients, seizure-specific characterization has to be initially performed within each subject. Identifying seizure-specific power changes and seizure-characteristic frequency band(s) is a key step of this endeavor and the main focus of this paper. Methods proposed in this paper provide a systematic tool for this key step. As shown in the simulations, the proposed direct GML and GACV methods are more robust and efficient than existing methods.

Spectral analysis is an important field in time series analysis [Brillinger (1981), Shumway (2000)]. The spectrum is often used to describe the power distribution and stochastic variation of a time series. For stationary time series, harmonic trends and power evolution can be explained by the spectrum at different frequencies. It is well known that stationarity does not always hold for EEG time series [Lopes da Silva (1978)]. Locally stationary processes have been proposed to approximate the nonstationary time series [Dahlhaus (1997), Ombao et al. (2002), Guo et al. (2003)]. The time-varying spectrum of a locally stationary time series characterizes changes of the stochastic variation over time that may reflect important patterns of biological activity. In particular, our spectral analysis of the EEGs can be used to aid the clinical judgement in two ways: first, due to the one-to-one correspondence between the spectrum and the variance of the EEG time series, changes in brain power before, during and after a seizure can be characterized by the estimated time-varying spectra of the pre-ictal and inter-ictal EEGs; second and more importantly, from a signal processing point of view, the phase information between two or more channels reveals whether the signals are synchronized, and at what frequencies (or frequency bands) if they are. This information provides an important marker of seizure initiation and localization [Mormann et al. (2003)]. By comparing the pre-seizure time-varying spectrum with that of the baseline (seizure-free) segments within a channel, the channel-specific seizure-characteristic frequencies can be identified (if such characteristics exist). If the seizure-specific frequency band(s) in one channel overlaps with that in another channel, then there will likely be the signal coupling between the two locations. Resecting either of the location would help to reduce surgical failure. However if the frequency band(s) do not overlap, then there may be across-frequency interactions that associated with the seizure propagation patterns [Mormann et al. (2000)].

As in most epileptic EEG studies, we analyze the intracranial EEGs (IEEGs) as the recordings have less artifacts. Figure 1 shows 5-minute IEEG segments from a patient with medicine-resistant mesial temporal lobe epilepsy right before a seizure's clinical onset (upper panel) and at baseline (lower panel) extracted at least four hours before the seizure's onset. The data were collected by the EEG Lab of University of Pennsylvania [D'Alessandro et al. (2001)]. It is visually unclear what the seizure-specific frequency band and its time-varying patterns are for both channels. We will describe analyses of this data in Section 5.

It is well known that the periodograms is an unbiased, but not consistent, estimator of the spectrum. Therefore, periodogram smoothing is a popular tool for nonparametric spectra estimation. Smoothing techniques including kernel smoother [Lee (1997), Ombao et al. (2001), Hannig and Lee (2004)], smoothing spline [Wahba (1980), Pawitan and O'Sullivan (1994), Guo et al. (2003)], regression spline [Kooperberg, Stone and Truong (1995)], local polynomial [Fan and Kreutzberger (1998)] and wavelet [Gao (1997)] have been applied to obtain consistent estimators. One may smooth periodograms directly [Lee (1997), Ombao et al. (2001), Hannig and Lee (2004)], smooth logarithmic periodograms [Wahba (1980), Guo et al. (2003)], or use Whittle likelihood [Pawitan and O'Sullivan (1994), Guo et al. (2003)]. In this paper, we consider smoothing spline estimation based on the Whittle likelihood.

It is well-recognized that the choice of smoothing parameters is crucial to the performance of the smoothing methods. Many methods have been developed for kernel smoother [Lee (1997), Ombao et al. (2001), Hannig and Lee (2004)] and smoothing spline [Wahba (1980), Pawitan and O'Sullivan (1994)]. Generalized cross-validation (GCV), generalized maximum likelihood (GML) and unbiased risk (UBR) criteria can be used to select the smoothing parameter when fitting smoothing spline models on the logarithms scale [Wahba (1990)]. The logarithmic transformation is the first order approximation which is less efficient than the Whittle likelihood [Pawitan and O'Sullivan (1994), Fan and Kreutzberger (1998), Guo et al. (2003)]. For fitting smoothing splines using penalized Whittle likelihood, Pawitan and O'Sullivan (1994) developed a criterion based on an estimate of the risk function for the selection of the smoothing parameter. Guo et al. (2003) used an indirect GML method to select the smoothing parameter. The indirect approach does not guarantee convergence and may have inferior performance than direct methods (Section 4).

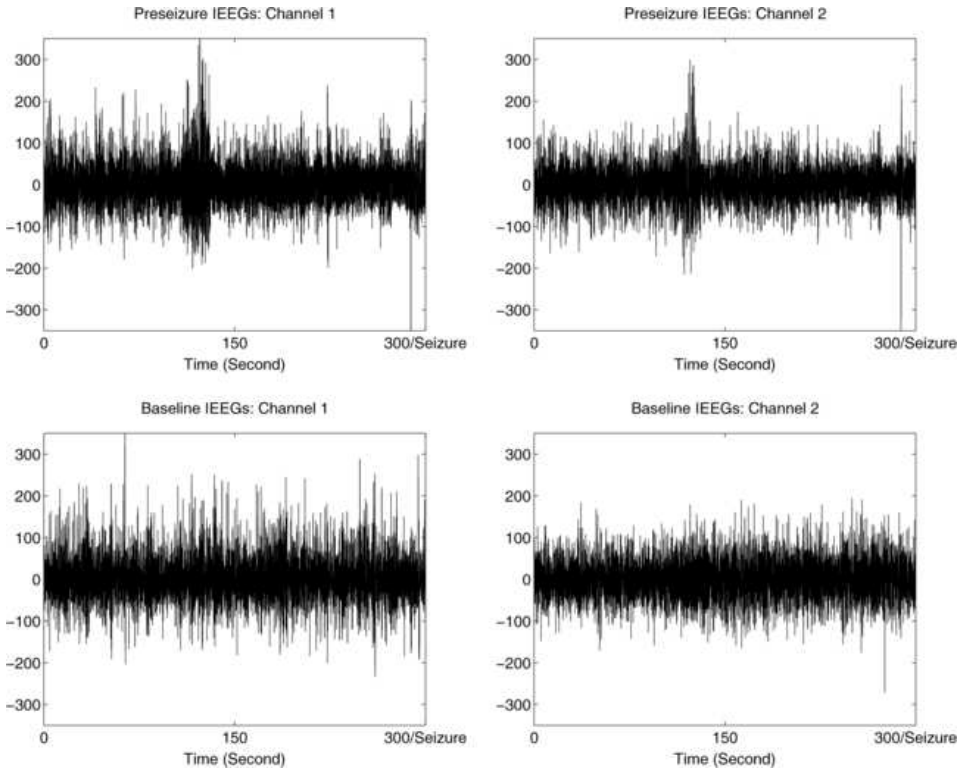

FIG. 1. *IEEG segments from two channels of an epileptic patient. The upper panels show* 5 *minute preseizure segments with the seizure's onset at the* 5*th minute. The lower panels show* 5 *minute baseline segments collected hours away from the seizure. The sampling rate is* 200 *Hertz. The total number of time points is 60,000 for each segment.*

We develop new direct data-driven methods for selecting smoothing parameters in the estimation of a spectral density and tests for the hypothesis that a locally stationary process is stationary. The rest of the paper is structured as follows. We present direct GML and GACV methods for stationary processes in Section 2. We develop direct GML method, GACV method and stationarity tests for locally stationary processes in Section 3. The simulation study is summarized in Section 4. The analysis of IEEG time series is presented in Section 5. We conclude with some remarks in Section 6.

## 2. Stationary processes.

2.1. *Notation.* Let $X_t$, $t = 0, \pm 1, \pm 2, \ldots$, be a stationary time series with mean zero and covariance function $\gamma(u) = \mathrm{E}(X_t X_{t+u})$. The second-order properties of $X_t$ are equivalently described by the spectrum

$$
(1) \qquad f(\omega) = \sum_{u=-\infty}^{\infty} \gamma(u) \exp(-i 2\pi \omega u), \qquad \omega \in [0, 1],
$$

where the imaginary unit $i^2 = -1$.

Let $X_0, X_1, \ldots, X_{T-1}$ be a finite sample of the stationary process and

$$
(2) \qquad y_k = T^{-1} \left| \sum_{t=0}^{T-1} X_t \exp(i 2\pi k t / T) \right|^2, \qquad k = 0, \ldots, T-1,
$$

be the periodogram at frequency $\omega_k = k/T$. Our goal is to estimate $f$ based on observations in the form $\{(\omega_k, y_k), k = 0, \ldots, T-1\}$. Under standard mixing conditions [Brillinger (1981)]

$$
(3) \qquad y_k \approx f(\omega_k) \chi^2_{k, d_k} / d_k,
$$

where $\chi^2_{k, d_k}$ are independent Chi-square random variables with degree of freedom $d_k = 1$ for $k = 0$ and $k = T/2$ (if $T$ is even) and $d_k = 2$ for $1 \leq k \leq T/2 - 1$. As [Pawitan and O'Sullivan (1994) and Guo et al. (2003)], for simplicity, the slight difference in degrees of freedom will be ignored and all degrees of freedom are set to two.

2.2. *Smoothing spline estimation.* We model the logarithm of the spectrum $g = \log(f)$ using a periodic spline space [Wahba (1990)]

$$
(4) \qquad \begin{aligned} W_2(\mathrm{per}) = \bigg\{ & g : g \text{ and } g' \text{ are abs. cont.}, \\ & g(0) = g(1), g'(0) = g'(1), \int_0^1 (g''(\omega))^2 \, d\omega < \infty \bigg\}. \end{aligned}
$$

$W_2(\mathrm{per}) = \{1\} \oplus W_2^0(\mathrm{per})$, where $\{1\}$ is the one-dimensional space consisting of all constant functions, and $W_2^0(\mathrm{per})$ is a reproducing kernel Hilbert space with reproducing kernel $R_1(\omega_1, \omega_2) = -B_4([\omega_1 - \omega_2])/24$, $[\omega_1 - \omega_1]$ is the fractional part of $\omega_1 - \omega_2$ and $B_4(\omega) = (\omega - 0.5)^4 - (\omega - 0.5)^2/2 + 7/240$ [Wahba (1990), Gu (2002)]. We note that [Wahba (1980) and Pawitan and O'Sullivan (1994)] essentially used the same model space with solutions approximated using cepstral coefficients.

Because $g(\omega)$ is symmetric about $\omega = 0.5$, it suffices to estimate $g(\omega)$ for $\omega \in [0, 0.5]$. As in Wahba (1980) and Pawitan and O'Sullivan (1994), we use all $y_k$'s even though $y_k = y_{T-k}$. This is to allow periodic smoothing. Let $g_k = g(\omega_k)$. Based on (3), we estimate $g$ as the minimizer of the following penalized Whittle likelihood [Gu (2002)]

$$\text{(5)} \qquad \sum_{k=0}^{T-1} \{g_k + y_k \exp(-g_k)\} + T\lambda \int_0^1 \{g''(\omega)\}^2 \, d\omega,$$

where $\lambda$ is a smoothing parameter balancing the goodness-of-fit and the smoothness of the function $g$.

For a fixed $\lambda$, the solution to (5) can be represented as [Wahba et al. (1995), Gu (2002)]

$$\text{(6)} \qquad \hat{g}(\omega) = d + \sum_{k=0}^{T-1} c_k R_1(\omega_k, \omega).$$

We compute coefficients $d$ and $c_k$'s iteratively using the iterative reweighted penalized least squares (IRPLS) method [McCullagh and Nelder (1989), Wahba et al. (1995)]. Specifically, at each iteration, we compute the *working variable* $z_k = \tilde{g}_k + y_k \exp(-\tilde{g}_k) - 1$ and *weight* $w_k = 1$ where tilde indicates current estimates. Note that, as [Pawitan and O'Sullivan (1994)], we use the Fisher scoring method instead of the Newton–Raphson method employed in Wahba et al. (1995), Gu (2002) and Liu, Tong and Wang (2006). Our experience indicates that the Fisher scoring method is more stable in this situation. One may select $\lambda$ using the GCV, GML or UBR criterion at each iteration of the above IRPLS procedure [Wahba et al. (1995)]. However, such an indirect approach may lead to nonconvergence of the algorithm (Section 4).

2.3. *Direct GML and GACV methods.* We simply introduce the direct GML and GACV criteria in this section. Derivations can be found in supplementary materials.

Consider the following prior for $g$:

$$\text{(7)} \qquad G(\omega) = \alpha + (T\lambda)^{-(1/2)} W(\omega),$$

where $\alpha \sim N(0, a)$, $W(\omega)$ is a Gaussian process independent of $\alpha$ with $E\{W(\omega)\} = 0$ and $E\{W(\omega_1)W(\omega_2)\} = R_1(\omega_1, \omega_2)$. [Gu (1992)] showed that, as $a \to \infty$, the posterior distribution of $G(\omega)$ can be approximated by a Gaussian distribution with posterior mean equals the spline estimate $\hat{g}$. This connection between smoothing splines and Bayesian models has been exploited to develop methods for selecting smoothing parameters for estimating variance functions [Liu, Tong and Wang (2006)] and constructing confidence intervals [Gu (1992), Wahba et al. (1995)]. We use this connection to develop a direct GML criterion for selecting $\lambda$.

Let $\mathbf{y} = (y_0, \ldots, y_{T-1})'$, $\mathbf{g} = (g_0, \ldots, g_{T-1})'$, $S = (1, \ldots, 1)$ be a vector of size $T$, $\Sigma = \{R_1(\omega_i, \omega_j)\}_{i,j=1}^T$, $(Q_1 Q_2)(R', \mathbf{0}')'$ be the QR decomposition of $S$, and $UDU'$ be the spectral decomposition of $Q_2'\Sigma Q_2$ where $D = \operatorname{diag}(\delta_1, \ldots, \delta_{T-1})$ and $\delta_\nu$ are eigenvalues. Let $\hat{g}_k = \hat{g}(\omega_k)$, $u_k = 1 - y_k \exp(-\hat{g}_k)$, $\hat{\mathbf{g}} = (\hat{g}_0, \ldots, \hat{g}_{T-1})'$, $\mathbf{u}_c = (u_0, \ldots, u_{T-1})'$, $\mathbf{y}_c = \hat{\mathbf{g}} - \mathbf{u}_c$ and $\mathbf{z} = (z_1, \ldots, z_{T-1})' = U'Q_2'\mathbf{y}_c$. Ignoring some constants, the negative log marginal likelihood of $\mathbf{y}$ can be approximated by

$$
\begin{aligned}
\operatorname{GML}(\lambda) = &\sum_{k=0}^{T-1} \{\hat{g}_k + y_k \exp(-\hat{g}_k)\} - \frac{1}{2}\mathbf{u}_c'\mathbf{u}_c \\
&+ \frac{1}{2}\sum_{\nu=1}^{T-1}\left\{\ln(\delta_\nu/T\lambda + 1) + \frac{z_\nu^2}{\delta_\nu/T\lambda + 1}\right\}.
\end{aligned}
\tag{8}
$$

The GML criterion (8) is new to the estimation of the spectrum. We minimize (8) to find an estimate of $\lambda$ which is referred to as the direct GML estimate.

As in Lin et al. (2000), consider the comparative Kullback–Leibler criterion

$$
\operatorname{CKL}(g, \hat{g}) = \frac{2}{T}\sum_{i=0}^{T-1}\{\exp(g_i - \hat{g}_i) + \hat{g}_i\}.
\tag{9}
$$

The CKL criterion cannot be used directly to select the smoothing parameters since it depends on the true log-spectrum. We need a proxy for this criterion. Let $\hat{g}^{(-i)}$ be the estimate of $g$ without the $i$th observation $y_i$, that is, $\hat{g}^{(-i)}$ is the minimizer of the penalized Whittle likelihood

$$
\sum_{k\neq i}\{g_k + y_k \exp(-g_k)\} + T\lambda \int_0^1 \{g''(\omega)\}^2 \, d\omega.
$$

Let $\hat{g}_k^{(-i)} = \hat{g}^{(-i)}(\omega_k)$. The leaving-out-one cross-validation estimate of the CKL criterion is

$$
\operatorname{CV}(\lambda) = \frac{2}{T}\sum_{i=0}^{T-1}\{y_i \exp(-\hat{g}_i^{(-i)}) + \hat{g}_i\}.
\tag{10}
$$

The function $\operatorname{CV}(\lambda)$ may be used to select the smoothing parameter. However, the computation is usually expensive. An approximation of $\operatorname{CV}(\lambda)$ is

$$
\begin{aligned}
\operatorname{GACV}(\lambda) = &\sum_{i=0}^{T-1}\{y_i \exp(-\hat{g}_i) + \hat{g}_i\} \\
&+ \frac{\operatorname{tr} H}{T - \operatorname{tr} W_0^{1/2} H W_0^{1/2}}\sum_{i=0}^{T-1} y_i \exp(-\hat{g}_i)\{y_i - \exp(\hat{g}_i)\},
\end{aligned}
\tag{11}
$$

where $H = (W + n\lambda\Omega)^{-1}V$, $W = \text{diag}(y_1 \exp(-\hat{g}_1)/2, \ldots, y_n \exp(-\hat{g}_n)/2)$, $\Omega = Q_2(Q_2'\Sigma Q_2)^+ Q_2'$, + represents the Moore–Penrose generalized inverse, $V = \text{diag}(\exp(-\hat{g}_1)/2, \ldots, \exp(-\hat{g}_n)/2)$ and $W_0 = \text{diag}(\exp(\hat{g}_1), \ldots, \exp(\hat{g}_n))$. Equation (11) is referred to as the GACV criterion and a GACV estimate of $\lambda$ is the minimizer of this criterion.

## 3. Locally stationary processes.

3.1. *Local periodograms.* Let $X_t, t = 0, \ldots, T-1$, be a finite sample of the following mean zero locally stationary process [Guo et al. (2003)]:

$$(12) \qquad X_t = \int_0^1 A(\omega, t/T) \exp(i2\pi\omega t) \, dZ(\omega),$$

where $Z(\omega)$ is a zero-mean stochastic process on [0, 1] and $A(\omega, u)$ denotes a transfer function with continuous second order derivatives for $(\omega, u) \in [0,1] \times [0,1]$.

The time-dependent spectrum $f(\omega, u) = \|A(\omega, u)\|^2$ is assumed to be a smooth function in both $\omega$ and $u$. For estimation, local periodograms are calculated on some pre-defined time-frequency grids. Specifically, the time domain is partitioned into $J$ disjoint blocks $[b_j, b_{j+1})$ where $0 = b_1 < b_2 < \cdots < b_J < b_{J+1} = 1$. Let $u_j = (b_j + b_{j+1})/2$ and $\omega_k$'s be $K$ frequencies in $[0, 1]$. Then the local periodograms are calculated as

$$(13) \qquad \hat{I}_{kj} = \hat{I}(\omega_k, u_j) = \frac{|\sum_{t=b_j}^{b_{j+1}-1} X_t \exp(i2\pi\omega_k t)|^2}{|b_{j+1} - b_j|},$$

$$k = 1, \ldots, K, \ j = 1, \ldots, J.$$

Guo et al. (2003) suggested that the block size should be in the order of $T^{1/2}$ and showed that the estimation is not very sensitive to the choices of sizes for the time and frequency grids. We investigate the impact of these choices in our simulations.

3.2. *SS ANOVA estimation.* As in Guo et al. (2003), we model logarithm of the time-dependent spectrum $g(w, u) = \log f(\omega, u)$ using the SS ANOVA model

$$(14) \quad g(\omega, u) = \beta_1 + \beta_2(u - 0.5) + s_1(\omega) + s_2(u) + s_3(\omega, u) + s_4(\omega, u),$$

where $\beta_2(u-0.5)$ and $s_2(u)$ are linear and smooth main effects of time, $s_1(\omega)$ is the smooth main effect of frequency, and $s_3(\omega, u)$ and $s_4(\omega, u)$ are linear-smooth and smooth-smooth interactions between frequency and time. Let $\boldsymbol{\gamma} = (\omega, u)$ and $\boldsymbol{\Gamma} = (\boldsymbol{\gamma}_i)_{i=1}^n$ be the selected time-frequency grid with corresponding log-local periodograms $\mathbf{y} = (y_1, \ldots, y_n) = (\log(\hat{I}_{11}), \ldots, \log(\hat{I}_{KJ}))$ where $n = KJ$. We estimate $g$ as the minimizer of the penalized Whittle likelihood

$$\text{(15)} \quad \sum_{i=1}^{n}\{g_i + y_i \exp(-g_i)\} + n\sum_{r=1}^{4}\lambda_r\|P_r g\|^2,$$

where $g_i = g(\boldsymbol{\gamma}_i)$, $\lambda_r$'s are smoothing parameters and $P_r$ is the projection operator onto the subspace corresponding to $s_r$, $r = 1,\ldots,4$. Let $\lambda_r = \lambda/\theta_r$. The solution to (15) is [Gu (2002)]

$$\text{(16)} \quad \hat{g}(\boldsymbol{\gamma}) = d_1 + d_2(u - 0.5) + \sum_{i=1}^{n} c_i \sum_{r=1}^{4} \theta_r R_r(\boldsymbol{\gamma}_i, \boldsymbol{\gamma}),$$

where $R_1(\omega_1, \omega_2)$ was defined in Section 2, $R_2(u_1, u_2) = B_2(u_1)B_2(u_2)/4 - B_4([\omega_1 - \omega_2])/24$, $R_3((\omega_1, u_1), (\omega_2, u_2)) = R_1(\omega_1, \omega_2)(u_1 - 0.5)(u_2 - 0.5)$, $R_4((\omega_1, u_1), (\omega_2, u_2)) = R_1(\omega_1, \omega_2)R_2(u_1, u_2)$, and $B_2(u) = (u - 0.5)^2 - 1/12$. Again, for fixed $\lambda_r$'s, coefficients $d_1$, $d_2$ and $c_i$'s can be computed using the IRPLS procedure. Guo et al. (2003) selected $\lambda_r$'s at each iteration using the GML method. Again, the indirect approach may lead to nonconvergence and inferior performance (Section 4).

3.3. *Direct GML and GACV methods.* We simply introduce the direct GML and GACV criteria in this section. Derivations can be found in supplementary materials.

Consider the following prior for $g$:

$$\text{(17)} \quad G(\boldsymbol{\gamma}) = \alpha_1 + \alpha_2(u - 0.5) + (n\lambda)^{-(1/2)} \sum_{r=1}^{4} \theta_r^{1/2} W_r(\boldsymbol{\gamma}),$$

where $\boldsymbol{\alpha} = (\alpha_1, \alpha_2)' \sim N(\boldsymbol{0}, aI)$, $W_r(\boldsymbol{\gamma})$'s are Gaussian processes independent of $\boldsymbol{\alpha}$ with $\mathrm{E}\{W_r(\boldsymbol{\gamma})\} = 0$ and $\mathrm{E}\{W_r(\boldsymbol{\gamma})W_r(\boldsymbol{\zeta})\} = R_r(\boldsymbol{\gamma}, \boldsymbol{\zeta})$.

Let $\mathbf{g} = (g_1, \ldots, g_n)'$, $S = \{1, u_i - 0.5\}_{i=1}^{n}$, $\Sigma_\theta = \{\sum_{r=1}^{4}\theta_r R_r(\boldsymbol{\gamma}_i, \boldsymbol{\gamma}_j)\}_{i,j=1}^{n}$, $(Q_1 Q_2)(R', \mathbf{0}')'$ be the QR decomposition of $S$, and $UDU'$ be the spectral decomposition of $Q_2'\Sigma_\theta Q_2$ where $D = \mathrm{diag}(\delta_1, \ldots, \delta_{n-2})$ and $\delta_\nu$ are eigenvalues. Let $\hat{g}_i = \hat{g}(\boldsymbol{\gamma}_i)$, $u_i = 1 - y_i \exp(-\hat{g}_i)$, $\hat{\mathbf{g}} = (\hat{g}_1, \ldots, \hat{g}_n)'$, $\mathbf{u}_c = (u_1, \ldots, u_n)'$, $\mathbf{y}_c = \hat{\mathbf{g}} - \mathbf{u}_c$, and $\mathbf{z} = (z_1, \ldots, z_{n-2})' = U'Q_2'\mathbf{y}_c$. Ignoring some constants, an approximation to the negative log marginal likelihood of $\mathbf{y}$ is

$$\text{(18)} \quad \mathrm{GML}(\lambda, \boldsymbol{\theta}) = \sum_{i=1}^{n}\{\hat{g}_i + y_i \exp(-\hat{g}_i)\} - \frac{1}{2}\mathbf{u}_c'\mathbf{u}_c \\ + \frac{1}{2}\sum_{\nu=1}^{n-2}\left\{\ln(\delta_\nu/n\lambda + 1) + \frac{z_\nu^2}{\delta_\nu/n\lambda + 1}\right\},$$

where $\boldsymbol{\theta} = (\theta_1, \ldots, \theta_4)'$. Note that $\delta_\nu$'s and $z_\nu$'s depend on $\boldsymbol{\theta}$. The direct GML estimates of the smoothing parameters $(\lambda, \boldsymbol{\theta})$ are the minimizers of (18).

The comparative Kullback–Leibler criterion and its cross-validation estimate are defined similarly as those in Section 2.3. Then an approximation to the cross-validation estimate of the CKL criterion is

$$\text{GACV}(\lambda, \boldsymbol{\theta}) = \sum_{i=1}^{n} \{y_i \exp(-\hat{g}_i) + \hat{g}_i\}$$
(19)
$$+ \frac{\operatorname{tr} H}{n - \operatorname{tr} W_0^{1/2} H W_0^{1/2}} \sum_{i=1}^{n} y_i \exp(-\hat{g}_i)\{y_i - \exp(\hat{g}_i)\},$$

where $H = (W + n\lambda\Omega)^{-1} V$, $W = \operatorname{diag}(y_1 \exp(-\hat{g}_1)/2, \ldots, y_n \exp(-\hat{g}_n)/2)$, $\Omega = Q_2(Q_2'\Sigma_\theta Q_2)^+ Q_2'$, $V = \operatorname{diag}(\exp(-\hat{g}_1)/2, \ldots, \exp(-\hat{g}_n)/2)$ and $W_0 = \operatorname{diag}(\exp(\hat{g}_1), \ldots, \exp(\hat{g}_n))$. The GACV estimates of the smoothing parameters $(\lambda, \boldsymbol{\theta})$ are the minimizer of (19).

3.4. *Permutation tests for stationarity.* It is often of interest to test if a locally stationary process is stationary:

$$H_0 : h(u) = 0 \quad \text{for all } u \quad \text{vs.} \quad H_1 : h(u) \neq 0 \quad \text{for some } u,$$

where $h(u) = \beta_2(u - 0.5) + s_2(u) + s_3(\omega, u) + s_4(\omega, u)$. The model under $H_1$ is the full SS ANOVA model (14). Denote the model under $H_0$, $g(\omega, u) = \beta_1 + s_1(\omega)$, as the reduced model. Let $\hat{g}^F$ and $\hat{g}^R$ be estimates of $g$ under the full and reduced models, respectively. Let $D_F = \sum_{i=1}^{n} \{\hat{g}_i^F + y_i \exp(-\hat{g}_i^F) - \log y_i - 1\}$ and $D_R = \sum_{i=1}^{n} \{\hat{g}_i^R + y_i \exp(-\hat{g}_i^R) - \log y_i - 1\}$ be deviances under the full and reduced models. We construct two statistics:

$$S_1 = D_R - D_F,$$
$$S_2 = \int_0^1 \int_0^1 \{\hat{g}^F(\omega, u) - \hat{g}^R(\omega, u)\}^2 \, d\omega \, du,$$

where $S_1$ corresponds to the Chi-square statistics commonly used for generalized linear models, and $S_2$ computes the $L_2$ distance between $\hat{g}^F$ and $\hat{g}^R$. We reject $H_0$ when these statistics are large. The null distributions of these statistics are unknown. Under $H_0$, $g$ does not depend on $u$. Therefore, we can use permutation to compute null distributions. Specifically, we generate permutation samples by shuffling time grid, compute two statistics for each permutation sample, and compute $p$-values as the proportion of statistics based on permutation samples which are larger than those based on the original data. Small scale simulations in the supplemental materials indicate that the permutation tests perform well.

## 4. Simulations.

4.1. *Simulations for stationary processes.* We first conduct simulations to evaluate the performance of the direct GML and GACV methods for spectral density estimation of stationary processes. We simulate data from two processes used in Wahba (1980) and Pawitan and O'Sullivan (1994):

1. AR(3): $X_t = 1.4256 X_{t-1} - 0.7344 X_{t-2} + 0.1296 X_{t-3} + \varepsilon_t$,
2. MA(4): $X_t = \varepsilon_t - 0.3\varepsilon_{t-1} - 0.6\varepsilon_{t-2} - 0.3\varepsilon_{t-3} + 0.6\varepsilon_{t-4}$,

where $\varepsilon_t \stackrel{\text{i.i.d.}}{\sim} N(0,1)$. We consider two sample sizes, $T = 128$ and $T = 256$, for each process. For each setting, we repeat simulation 1000 times.

To compare with the method in Wahba (1980), we fit cubic periodic splines to the logarithmic transformed periodograms $z_k = \log(y_k) + c_k$ where $c_k = 0.57721$ for $k \neq 0$ and $k \neq T/2$ and $c_k = 0.30135$ for $k = 0$ or $k = T/2$. We use the GML method to select the smoothing parameter on the logarithmic scale. Other methods such as GCV have also been tested and the comparative results remain the same. To compare with the method in Pawitan and O'Sullivan (1994), we fit a cubic periodic spline model using the penalized likelihood (5) with the smoothing parameter $\lambda$ selected as the minimizer of

$$RE(\lambda) = \sum_{k=0}^{T-1} (\hat{g}_k - v_k)^2 + 2\,\text{tr}\{H(\lambda)\}, \tag{20}$$

where $v_k = \hat{g}_k + y_k \exp(-\hat{g}_k) - 1$ and $H(\lambda)$ is the smoother matrix at convergence of the IRPLS procedure with the fixed $\lambda$. The criterion (20) is a direct adaption of (12) in Pawitan and O'Sullivan (1994). $v_k$'s need to be calculated with $\lambda$ close to the optimal choice. Pawitan and O'Sullivan (1994) suggested a two-step procedure: first choose $\lambda$ that minimizes $RE(\lambda)$ using $v_k = \hat{g}_k + y_k \exp(-\hat{g}_k) - 1$ and denote the resulting estimates as $\hat{g}_{kp}$, and then choose $\lambda$ that minimizes $RE(\lambda)$ using $v_k = \hat{g}_{kp} + y_k \exp(-\hat{g}_{kp}) - 1$. We use the same two-step procedure with a uniform (on a logarithmic scale) grid search over the range $e^{-25} \leq \lambda \leq e^{-1}$. As Pawitan and O'Sullivan (1994), 5 values are used in the first step and 50 values are used in the second step.

For each simulation replication, we compute the mean squared error as

$$\text{MSE}_m = \frac{1}{T} \sum_{k=0}^{T-1} (\hat{g}_k - g_k)^2,$$

where $m$ represents the method employed to estimate $g$. $m = $ LS, IM, DM, DV and PO, respectively, represent the method in Wahba (1980), the indirect GML method, the direct GML method, the GACV method and the method in Pawitan and O'Sullivan (1994). The indirect GML method selects $\lambda$ at each iteration of the IRPLS procedure using the GML criterion.

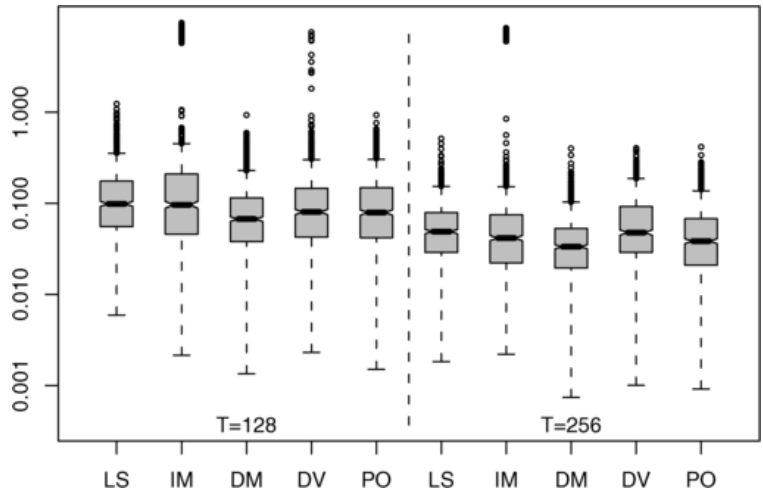

Fig. 2. *Boxplots of MSEs on logarithm scale for the AR3 process.*

For the indirect GML method, several replications failed to converge (Table 1). The GACV method failed to converge in one replication. None of the other methods has this problem. The boxplots of MSEs are shown in Figures 2 and 3. As Pawitan and O'Sullivan (1994), we compute the relative efficiencies $\text{MSE}_m/\text{MSE}_{\text{DM}}$ for $m = $ LS, IM, DV and PO. The medians of these relative efficiencies are listed in Table 1. From Figures 2 and 3, the MSEs based on the LS, IM, DV and PO methods have heavier tails than those from the DM method. Thus the mean relative efficiencies can be much bigger than medians (Table 1). We conclude that the direct GML is stable and has the best overall performance. Even though involving a somewhat ad hoc two-step procedure, the PO method performs well. One problem with the PO method is that it sometimes undersmooth the spec-

TABLE 1
*Median (first row for each $T$) and mean (second row for each $T$) relative efficiencies (relative to the DM method)*

|     | AR3  |          |      |      | MA4  |           |         |      |
|-----|------|----------|------|------|------|-----------|---------|------|
| $T$ | LS   | IM       | DV   | PO   | LS   | IM        | DV      | PO   |
| 128 | 1.51 | 1.22 (2) | 1.41 | 1.08 | 1.30 | 10.37 (7) | 1.77    | 1.03 |
|     | 1.82 | 20.06    | 3.70 | 1.45 | 1.41 | 10.83     | 2.29    | 1.12 |
| 256 | 1.49 | 1.19 (4) | 1.19 | 1.06 | 1.33 | 11.80 (11)| 1.09 (1)| 0.99 |
|     | 1.83 | 16.87    | 3.13 | 1.45 | 1.43 | 13.76     | 1.44    | 1.05 |

Numbers in the parentheses are the number of replications out of 1000 simulation replications that the indirect GML and GACV methods failed to converge.

trum. The LS method is less efficient. The indirect GML method sometimes fails to converge and performs very badly for the MA4 process. The GACV method occasionally fails to converge due to numerical problems. Its performance is inferior to the direct GML when it converges. This is especially true in the case when $n$ is small where GACV has many large MSEs due to undersmoothing, a phenomenon has been previously observed for the GCV method [Wahba and Wang (1993)]. We have conducted simulations with other stationary processes and sample sizes. The comparative results remain the same.

4.2. *Simulations for locally stationary processes.* To investigate the performance of the direct GML and GACV methods for locally stationary processes, and compare them with the LS and indirect GML methods, we simulate locally stationary time series from two time-varying spectra:

1. LS1: $g(\omega, u) = 4 + \sin(2\pi u) + \log\{1.25 - \cos(2\pi\omega)\}$.
2. LS2: $g(\omega, u) = 5 - 8(\omega - 0.5)^2 + \sin\{2\pi \exp(u)\} + 0.01(\omega - 0.5)^2 \times \sin\{2\pi \exp(u)\}$.

Similarly to [Guo and Dai (2006)], the time series $X_t, t = 0, \ldots, T - 1$, are simulated based on the following relationship:

$$(21) \qquad X_t = \sum_{k=0}^{T-1} \exp\{g(k/T, t/T)\} \exp(2\pi k t i/T) Z_k,$$

where $Z_k = Z(k/T)$ are mutually independent random variables distributed as complex Normal with mean zero and variance $1/T$. $Z_k = \overline{Z_{T-k}}$ when $k/T \neq 0, 0.5$, or 1. When $k/T = 0, 0.5$, or 1, $Z_k = Z(k/T)$ are mutually independent random variables distributed as real Normal with mean zero and variance $1/T$. We consider two sample sizes, $T = 1024$ and $T = 2048$, for each locally stationary process. To assess the impact of block sizes, we consider three time-frequency grids: $(K, J) = (64, 16)$, $(K, J) = (32, 32)$ and $(K, J) = (16, 64)$. For each setting, we repeat simulation 100 times.

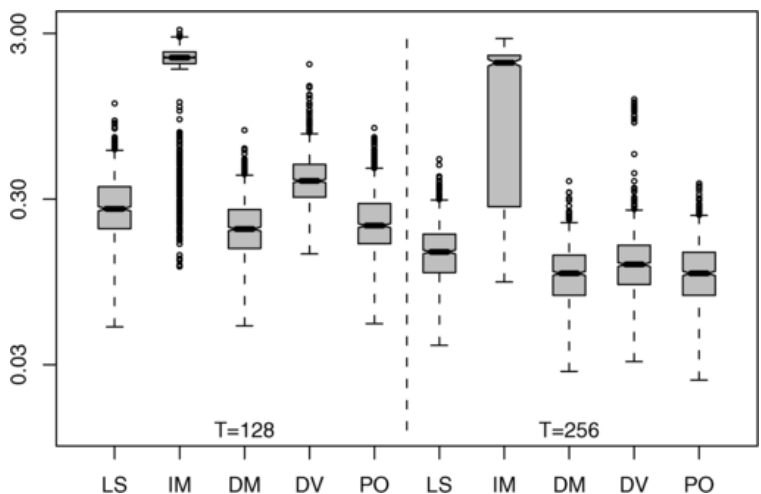

FIG. 3. *Boxplots of MSEs on logarithm scale for the MA4 process.*

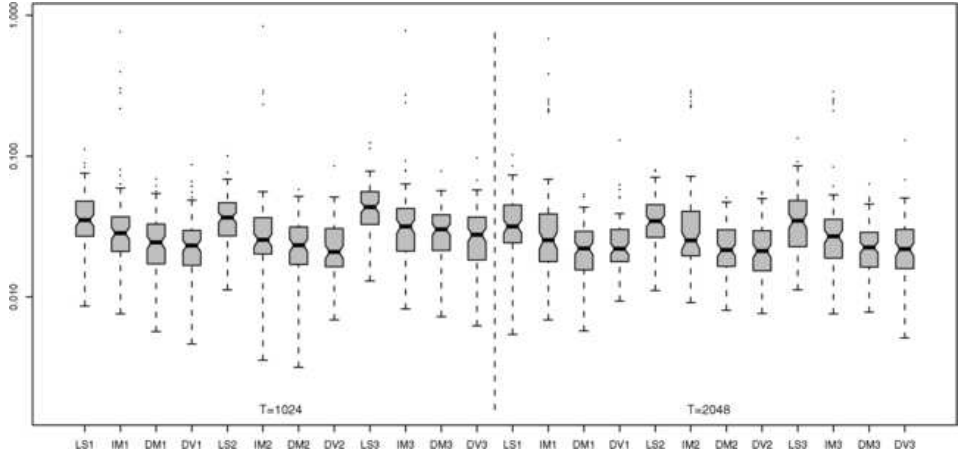

FIG. 4. *Boxplots of MSEs on logarithm scale for the LS1 process. Numbers in the x-axis labels resent three settings of time-frequency grids:* 1 *for* $(64, 16), 2$ *for* $(32, 32)$ *and* 3 *for* $(16, 64)$.

We compute the mean squared error as

$$\mathrm{MSE}_m = \sum_{k=1}^{K} \sum_{j=1}^{J} (\hat{g}_{kj} - g_{kj})^2 / (KJ),$$

where $m =$ LS, IM, DM and DV which correspond to the method based on logarithmic transformation, the indirect GML method in [Guo et al. (2003)] and the direct GML method and the GACV method.

The boxplots of MSEs are shown in Figures 4 and 5. Table 2 lists median relative efficiencies and the number of replications that the indirect GML and GACV methods failed to converge. The comparative results are similar to those in the stationary case: the indirect GML and GACV methods may fail to converge and the direct GML is stable and has the best performance. When converged, the GACV has comparable performance to the direct GML method. However, the GACV method takes much longer to compute. Therefore, the direct GML method is recommended. The estimation is not very sensitive to the choices of block sizes. We have conducted simulations with other sample sizes and block sizes. The comparative results remain the same.

TABLE 2
*Median relative efficiencies (relative to the DM method)*

| | | LS1 | | | LS2 | | |
|---|---|---|---|---|---|---|---|
| $T$ | $(K, J)$ | LS | IM | DV | LS | IM | DV |
| 1024 | (64, 16) | 1.49 | 1.14 (4) | 0.98 | 1.49 | 1.27 (13) | 1.02 (1) |
| | (32, 32) | 1.55 | 1.10 (6) | 0.91 | 1.38 | 1.22 (13) | 1.06 (1) |
| | (16, 64) | 1.47 | 1.14 (11) | 0.99 | 1.39 | 1.21 (10) | 1.18 |
| 2048 | (64, 16) | 1.48 | 1.17 (6) | 1.05 | 1.50 | 1.33 (12) | 1.07 (1) |
| | (32, 32) | 1.51 | 1.13 (5) | 1.12 (1) | 1.53 | 1.28 (8) | 1.18 (1) |
| | (16, 64) | 1.56 | 1.19 (3) | 0.95 (1) | 1.53 | 1.16 (10) | 1.26 (1) |

Numbers in parentheses are the number of replications out of 100 simulation replications that the indirect GML and GACV methods failed to converge.

**5. The IEEG analysis.** Figure 1 shows the IEEGs of 5-minute preseizure and baseline segments from two channels. These two channels recorded IEEGs from adjacent electrodes in the mesial temporal lobe of the brain believed to be most relevant to the seizure. Therefore, we expect to see close connections in the power and frequency activities between both channels. The important clinical questions are: (1) does the characteristic frequencies change over time during the pre-ictal stage of the seizure as compared to the baseline in each channel? and (2) are such frequencies common across both channels? We answer these questions by analyzing the IEEGs using our estimation procedure.

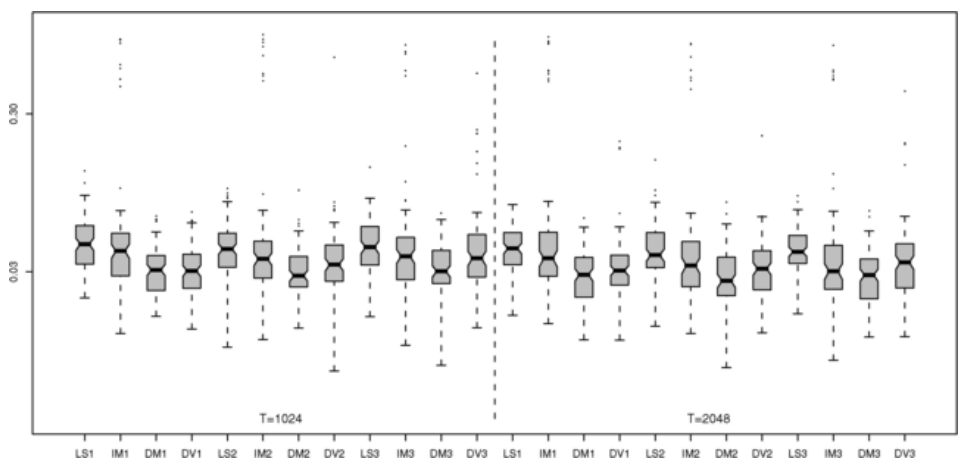

FIG. 5. *Boxplots of MSEs on logarithm scale for the LS2 process. Numbers in the x-axis labels resent three settings of time-frequency grids:* 1 *for* (64, 16), 2 *for* (32, 32) *and* 3 *for* (16, 64).

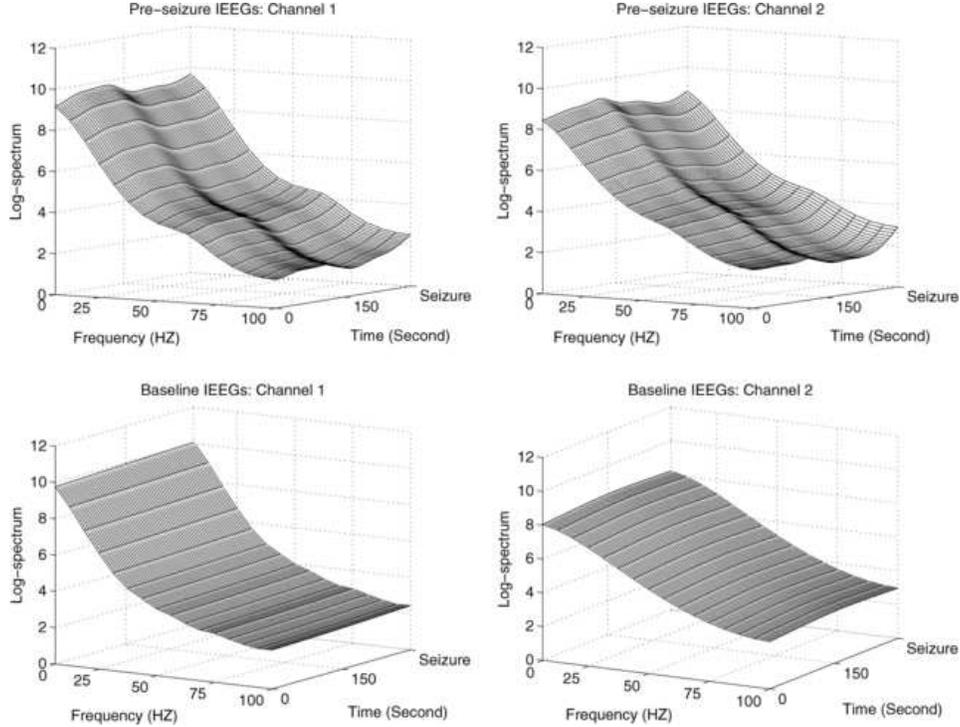

Fig. 6. *Time-varying spectra estimates (in log scale) for the preseizure segments (upper panels) and the baseline segments (lower panels) of the 5-minute IEEG segments from channel 1 (left) and channel 2 (right).*

We assume that the IEEG segments are locally stationary. To remove artifacts, we first normalize the raw IEEG recordings by subtracting their across-channel means. Each segment is then partitioned into 64 time blocks. Thirty-two equally-spaced frequency points are selected to calculate the local periodograms. Specifically, the time-frequency grids for the calculation are $(\omega_k, u_j) = (k/33, (938 \times j - 468.5)/60000)$ for $k = 1, \ldots, 32, j = 1, \ldots, 63$ and $(\omega_k, u_{64}) = (k/33, 0.9925)$ for $k = 1, \ldots, 32$. Figure 6 shows the SS ANOVA estimates of time-varying spectra for both the baseline and preseizure segments of the two channels. The direct GML method is used for all fits in this section. Because the sampling rate is 200 HZ and the spectrum is symmetric around 100 HZ, we can only assess power changes in frequency bands ([0 HZ–100 HZ]). It appears that the baseline spectrum does not vary much over time.

For channel 1, the $p$-values for testing stationarity based on two statistics are 0.81 and 0.73 for the baseline segment and 0.01 and 0.01 for the preseizure segment. For channel 2, the $p$-values are 0.31 and 0.16 for the baseline segment and 0 and 0.0067 for the preseizure segment. We conclude that for this data set, the processes far away from the seizure's clinical onset can be regarded as stationary while the processes close to the seizure's clinical onset is nonstationary. As expected the pre-ictal spectra have similar shapes, indicating the possible connectivity between the power evolutions of the two adjacent channels.

To find significant changes of the preseizure power spectra from those of the baseline segments, we compute 95% Bayesian confidence intervals for the estimated preseizure power spectra using the same method described in Wahba et al. (1995). We also compute the difference in estimated power spectra between preseizure and baseline IEEGs. At a particular point of time and frequency, the difference is deemed significant if the estimated baseline power spectrum is outside the 95% confidence interval. Figure 7 shows the contour plots of estimated power spectra differences for time-frequency regions where the differences are significant.

In channel 1, a high-frequency power discharge (at [75 HZ–100 HZ]) was recorded during approximately $-300$ to $-280$ seconds before the seizure; then a power build-up was captured for the next 85 seconds at [10 HZ–40 HZ], followed by another significant power decrease at both high and low frequency ends ([75 HZ–100 HZ] and $<5$ HZ). However, the channel 2 IEEGs recorded significant power discharges as a broad band activity ([75 HZ–100 HZ] and [20 HZ–40 HZ]) 5 minutes before the seizure. Interestingly, in this channel power increased around the same time as channel 1 ($-270$ to $-150$ seconds), but at lower frequencies ($<10$ HZ). This phenomenon implies that the short term power build up may be regarded as a warning signal of the coming seizure.

For both channels, the common frequency band for decreased power is within [75 HZ–100 HZ]. While there are no common frequencies at which the power increase occurred, the across-frequency interaction between the two channels appeared during $-270$ to $-150$ seconds. By further studying the within-frequency (at [75 HZ–100 HZ]) and across-frequency (between [10 HZ–40 HZ] and [0 HZ–10 HZ]) signal coupling, the seizure origination and propagation path may be inferred, which would eventually allow preventive action or surgical resection to take place at the right location to prevent a future seizure. Our results conform with recent findings in the literature. Specifically, high-frequency oscillations are usual suspects of the electrical activities for the epileptic brain, which may be important markers for epileptic network function ([80 HZ–500 HZ]) [Worrell et al. (2004)], and $\gamma$-band ([25 HZ–60 HZ]) activity may be associated with a reliable warning of the seizure [Aksenova, Volkovych and Villa (2007)].

**6. Discussion.** We have developed methods and software for smoothing spline and SS ANOVA spectrum estimation of stationary and locally stationary processes. These methods are recommended since they are stable and perform better than other existing methods. The risk estimation method by Pawitan and O'Sullivan (1994) may also be extended to the SS ANOVA spectrum estimation of locally stationary processes. It is unlikely that such an extension will perform better than the direct GML method. For a stationary process, similar permutation tests can be constructed to test if its spectrum is a constant which amounts to a white noise series [Fan and Zhang (2004)].

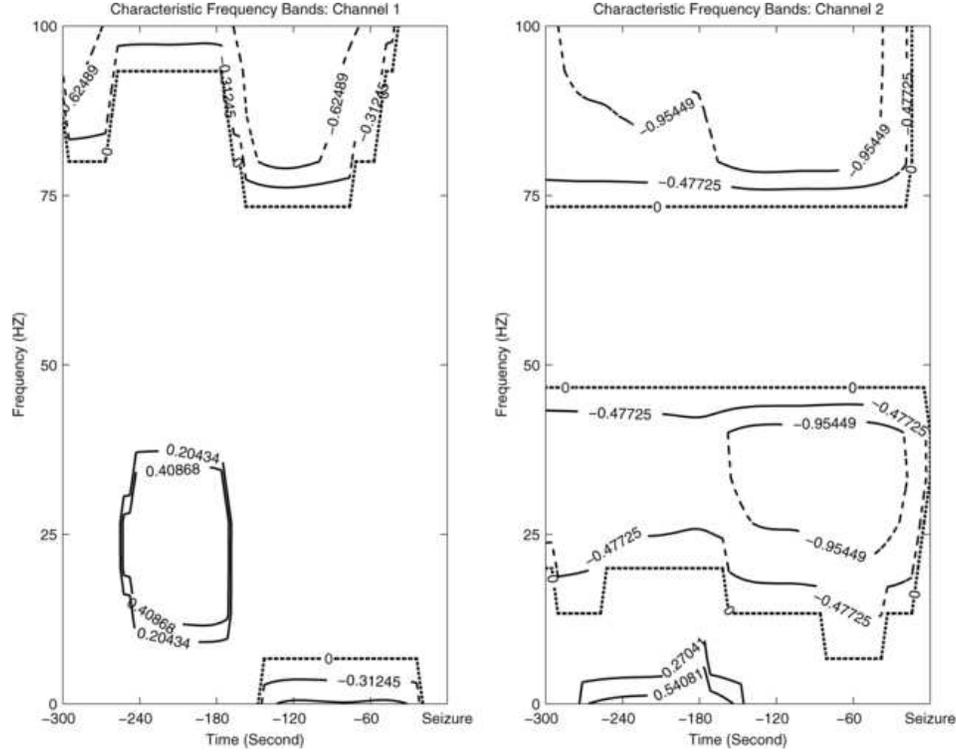

FIG. 7. *Contour plots for the significant differences in power spectra between the preseizure IEEGs and baseline IEEGs (defined by subtracting the spectra estimates of the baseline segments from those of the preseizure segments). The positive differences (solid lines) and negative differences (dashed lines) indicate significant increases and significant decreases in power at 5% significance level. For both channels, the common frequency band for power discharge is within [75 HZ–100 HZ], occurred as early as 5 minutes before its onset, while the simultaneous power build-up between $-270$ and $-150$ seconds happened at different frequencies ([10 HZ–40 HZ] for channel 1 and $<10$ HZ for channel 2).*

Our analyses have successfully picked up the important time-varying behaviors of the power and frequency components of the IEEG channels, which may be indicative of the seizure onset and propagation patterns [Worrell et al. (2004), Mormann et al. (2003), Mormann et al. (2000), Schiller, Cascino, Busacker and Sharbrough (1998)]. Future analyses would include adopting the technique for multiple channels and multiple seizures. With the aid of clinical judgements from neurologists and neurosurgeons, seizure propagation patterns may eventually be uncovered.

Even though we emphasize the application to the spectral analysis of EEG time series, our methods are general with a wide range of applications including neurological cognition studies. And they may be applied to understand the periodic secretion patterns of hormone time series and assess the spectral properties of the magnetoencephalography (MEG) time series.

**Acknowledgments.** The authors thank the Editor, Associate Editor and two references for constructive comments that substantially improved an earlier draft.

## SUPPLEMENTARY MATERIAL

**Derivation of the direct GML criterion permutation test of stationarity: SOM for nonparametric spectral analysis** (DOI: 10.1214/08-AOAS185SUPP; .pdf). The supplementary material [see Qin and Wang (2008)] contains derivation of the direct GML criterion, derivation of the GACV criterion, simulations for permutation tests of stationarity and R code for spectrum estimation.

## REFERENCES

AKSENOVA, T., VOLKOVYCH, V. and VILLA, A. (2007). Detection of spectral instability in EEG recordings during the preictal period. *J. Neural Engineering* **4** 173–178.

BRILLINGER, D. (1981). *Time Series: Data Analysis and Theory*, 2nd ed. Holden Day, San Francisco. MR0595684

DAHLHAUS, R. (1997). Fitting time series models to nonstationary processes. *Ann. Statist.* **25** 1–37. MR1429916

D'ALESSANDRO, M., VACHTSEVANOS, G., ESTELLER, R., ECHAUZ, J. and LITT, B. (2001). A generic approach to selecting the optimal feature for epileptic seizure prediction. *IEEE International Meeting of the Engineering in Medicine and Biology Society*.

FAN, J. and KREUTZBERGER, E. (1998). Automatic local smoothing for spectral density estimation. *Scand. J. Statist.* **25** 359–369. MR1649039

FAN, J. and ZHANG, W. (2004). Generalised likelihood ratio tests for spectral density. *Biometrika* **91** 195–209. MR2050469

GAO, H.-Y. (1997). Choice of thresholds for wavelet shrinkage estimate of the spectrum. *J. Time Ser. Anal.* **18** 231–251. MR1456639

GU, C. (1992). Penalized likelihood regression: A Bayesian analysis. *Statist. Sinica* **2** 255–264. MR1152308

GU, C. (2002). *Smoothing Spline ANOVA Models*. Springer, New York. MR1876599

GUO, W. and DAI, M. (2006). Multivariate time-dependent spectral analysis using cholesky decomposition. *Statist. Sinica* **16** 825–845. MR2281304

GUO, W., DAI, M., OMBAO, H. C. and VON SACHS, R. (2003). Smoothing spline ANOVA for time-dependent spectral analysis. *J. Amer. Statist. Assoc.* **98** 643–652. MR2011677

HANNIG, J. and LEE, T. (2004). Kernel smoothing of periodograms under Kullback–Leibler discrepancy. *Signal Process.* **84** 1255–1266.

KOOPERBERG, C., STONE, C. J. and TRUONG, Y. K. (1995). Logspline estimation of a possibly mixed spectral distribution. *J. Time Ser. Anal.* **16** 359–388. MR1342682

LEE, T. (1997). A simple span selector for periodogram smoothing. *Biometrika* **84** 965–969. MR1625012

LIN, X., WAHBA, G., XIANG, D., GAO, F., KLEIN, R. and KLEIN, B. (2000). Smoothing spline ANOVA models for large data sets with Bernoulli observations and the randomized GACV. *Ann. Statist.* **28** 1570–1600. MR1835032

LIU, A., TONG, T. and WANG, Y. (2006). Smoothing spline estimation of variance functions. *J. Comput. Graph. Statist.* **16** 312–329. MR2370945

LOPES DA SILVA, F. (1978). Analysis of EEG nonstationarities. In *Contemporary Clinical Neurophysiology* (W. A. Cobb and H. van Dujn, eds.) 165–179. Oxford Univ. Press.

MCCULLAGH, P. and NELDER, J. (1989). *Generalized Linear Models*. Chapman and Hall, London. MR0727836

MORMANN, F., ANDRZEJAK, R., KREUTZ, T., RIEKE, C., DAVID, P., ELGER, C. and LEHNERTZ, K. (2003). Automated detection of a preseizure state based on a decrease in synchronization in intracranial electroencephalogram recordings from epilepsy patients. *Phys. Rev. E* **67** 1–10.

MORMANN, F., LEHNERTZ, K., DAVID, P. and ELGER, C. (2000). Mean phase coherence as a measure for phase synchronization and its application to the EEG of epilepsy patients. *Phys. D* **144** 358–369.

OMBAO, H. C., RAZ, J. A., STRAWDERMAN, R. L. and VON SACHS, R. (2001). A simple generalised crossvalidation method of span selection for periodogram smoothing. *Biometrika* **88** 1186–1192. MR1872229

OMBAO, H., RAZ, J., VON SACHS, R. and GUO, W. (2002). The SLEX model of a nonstationary time series. *Ann. Inst. Statist. Math.* **54** 171–200. MR1893549

PAWITAN, Y. and O'SULLIVAN, F. (1994). Nonparametric spectral density estimation using penalized Whittle likelihood. *J. Amer. Statist. Assoc.* **89** 600–610. MR1294086

QIN, L. and WANG, Y. (2008). Supplement to "Nonparametric spectral analysis with applications to seizure characterization using EEG time series." DOI: 10.1214/08-AOAS185SUPP.

SCHILLER, Y., CASCINO, G. D., BUSACKER, N. E. and SHARBROUGH, F. W. (1998). Characterization and comparison of local onset and remote propagated electrographic seizures recorded with intracranial electrodes. *Epilepsia* **39** 380–388.

SCHILLER, Y., CASCINO, G. D. and SHARBROUGH, F. W. (1998). Chronic intracranial EEG monitoring for localizing the epileptogenic zone: An electroclinical correlation. *Epilepsia* **39** 1302–1308.

SHUMWAY, R. H. and STOFFER, D. S. (2000). *Time Series Analysis and Its Applications*. Springer, New York. MR1856572

WAHBA, G. (1980). Automatic smoothing of the log periodogram. *J. Amer. Statist. Assoc.* **75** 122–132.

WAHBA, G. (1990). *Spline Models for Observational Data*. SIAM, Philadelphia. MR1045442

WAHBA, G. and WANG, Y. (1993). Behavior near zero of the distribution of GCV smoothing parameter estimates for splines. *Statist. Probab. Letters* **25** 105–111. MR1365026

WAHBA, G., WANG, Y., GU, C., KLEIN, R. and KLEIN, B. (1995). Smoothing spline ANOVA for exponential families, with application to the Wisconsin Epidemiological Study of Diabetic Retinopathy. *Ann. Statist.* **23** 1865–1895. MR1389856

WINTERHALDER, M., TAIWALD, T., VOSS, H. U., ASCHENBRENNER-SCHEIBE, R., TIMMER, J. and SCHULZE-BONHAGE, A. (2003). The seizure prediction characteristic: A general framework to assess and compare seizure prediction methods. *Epilepsy and Behavior* **4** 318–325.

WORRELL, G. A., PARISH, L., CRANSTOUN, S. D., JONAS, R., BALTUCH, G. and LITT, B. (2004). High-frequency oscillations and seizure generation in neocortical epilepsy. *Brain* **127** 1496–1506.

PROGRAM IN BIOSTATISTICS
VACCINE AND INFECTIOUS DISEASES INSTITUTE
DIVISION OF PUBLIC HEALTH SCIENCES
FRED HUTCHINSON CANCER RESEARCH CENTER
SEATTLE, WASHINGTON 98109
USA
E-MAIL: lqin@fhcrc.org

DEPARTMENT OF STATISTICS
AND APPLIED PROBABILITY
UNIVERSITY OF CALIFORNIA
SANTA BARBARA, CALIFORNIA 93106
USA
E-MAIL: yuedong@pstat.ucsb.edu